\documentclass[iop,apj,tighten]{emulateapj}
\usepackage{apjfonts} 
\usepackage{amsmath,amstext}
\usepackage[breaklinks,colorlinks,citecolor=blue,linkcolor=magenta]{hyperref} 
\usepackage[all]{hypcap} 
\usepackage{threeparttable}
\usepackage{bm}
\usepackage{ulem}

\newcommand{\kms}{km$\rm s^{-1}$}
\newcommand{\masyr}{\,mas\,$\rm yr^{-1}$}
\newcommand{\mura}{$\mu_{\alpha^*}$}
\newcommand{\mudec}{$\mu_{\delta}$}

\newcommand{\mmura}{$\langle \epsilon_{\mu_{\alpha^*}}\rangle$}
\newcommand{\mmudec}{$\langle \epsilon_{\mu_{\delta}}\rangle$}

\newcommand{\dmura}{$\epsilon_{\mu_{\alpha^*}}$}
\newcommand{\dmudec}{$\epsilon_{\mu_{\delta}}$}

\newcommand{\HJT}[1]{{{\color{black} #1}}}

\newcommand{\HJ}[1]{{{\color{red}{\bf } #1}}}

\shorttitle{GPS1+ proper motions}
\shortauthors{HJT, et al.}

\begin{document}

\title{The extended Gaia-PS1-SDSS (GPS1+) proper motion catalog}

\author{
Hai-Jun Tian\altaffilmark{1,2}, 
Yang Xu\altaffilmark{3},
Chao Liu\altaffilmark{3},
Hans-Walter Rix\altaffilmark{2},
Branimir Sesar\altaffilmark{2}
Bertrand Goldman\altaffilmark{2,4}
}

\altaffiltext{1}{Center for Astronomy and Space Sciences, China Three Gorges University, Yichang 443002, China.}
\altaffiltext{2}{Max Planck Institute for Astronomy, K\"onigstuhl 17, D-69117 Heidelberg, Germany. Email: hjtian@lamost.org}
\altaffiltext{3}{Key Lab for Optical Astronomy, National Astronomical Observatories, Chinese Academy of Sciences, Beijing 100012, China.}
\altaffiltext{4}{Observatoire astronomique de Strasbourg, Universit\'e de Strasbourg, CNRS, UMR 7550, 11 rue de l'Universit\'e, F-67000 Strasbourg, France.}

\begin{abstract}
The GPS1 catalog was released in 2017. It delivered precise proper motions for around 350 million sources across three-fourths of the sky down to a magnitude of $r\sim20$\,mag. In this study, we present GPS1+  the extension GPS1 catalog down to $r\sim22.5$\,mag, based on {\it Gaia} DR2, PS1, SDSS and 2MASS astrometry. The GPS1+ totally provides proper motions for $\sim$400 million sources with a characteristic systematic error of less than 0.1\masyr. This catalog is divided into two sub-samples, i.e., the primary and secondary parts. The primary $\sim$264 million sources have either or both of the {\it Gaia} and SDSS astrometry, with a typical precision of 2.0-5.0 \masyr. In this part, $\sim$160 million sources have {\it Gaia} proper motions, we provide another new proper motion for each of them by building a Bayesian model. Relative to {\it Gaia}'s values, the precision is improved by $\sim$0.1\,dex on average at the faint end; $\sim$50 million sources are the objects whose proper motions are missing in {\it Gaia} DR2, we provide their proper motion with a precision of $\sim$4.5\masyr; the remaining $\sim$54 million faint sources are beyond {\it Gaia} detecting capability, we provide their proper motions for the first time with a precision of 7.0 \masyr. However, the secondary $\sim$136 million sources only have PS1 astrometry, the average precision is worse than 15.0 \masyr. All the proper motions have been validated using QSOs and the existing {\it Gaia} proper motions. The catalog will be released on-line and available via the TAP Service, \HJ{or via the National Astronomical Data Center serviced by China-VO: https://nadc.china-vo.org/data/data/gps1p/f}

 \end{abstract}

\section{Introduction}\label{sect:intro}
{\it Gaia}, a cornerstone mission of the European Space Agency (ESA), is ambitious to chart a three-dimensional map of our Galaxy with unprecedented precision. After 22 months of observations, {\it Gaia} delivered its second release ({\it Gaia} DR2) on April 25, 2018 \citep{gaia2018}. This catalog contains the positions of nearly 1.7 billion objects with G-band magnitude brighter than $\sim$ 20.7. Among these sources, more than 1.3 billion stars  in the Milk Way have precise positions, proper motions, parallaxes and colors. The average uncertainties in the respective proper motion components are up to 0.06 \masyr (for $G<15$\,mag), 0.2 \masyr (for $G=17$\,mag) and 1.2 \masyr (for $G = 20$\,mag). The {\it Gaia} DR2 parallaxes and proper motions are based only on {\it Gaia} data.

The {\it Gaia} DR2 supersedes the most majority of current existing proper motion catalogs. The previous proper motion catalogs, such as PPMXL \citep{roeser2010}, HSOY \citep{Altmann2017}, the UCAC series \citep{zacharias2004, zacharias2010, Zacharias2017}, APOP \citep{qi2015}, and GPS1 \citep[][hereafter, T17]{tian2017}, are not comparable with {\it Gaia} DR2 in quality, even though HSOY, UCAC5 and GPS1 were built combining the precise {\it Gaia} DR1 astrometry \citep{gaia2016a}. 

Unfortunately, some limitations still exists in {\it Gaia} DR2: (1) more than 361 million sources only have positions (precision $\sim$2\,mas) at J2015.5 and the mean {\it G} magnitude, missing proper motions and parallax etc; (2) the average precision of proper motions is hard to reach a level of \,sub-mas\,$\rm yr^{-1}$ for faint sources, in particular for those close to the {\it Gaia} limiting magnitude; (3) {\it Gaia} DR2 is complete in $12 < G < 17$\,mag, but incomplete at an ill-defined faint magnitude limit; (4) no sources with $G > 20.7$\,mag. 

In this study, we would like to extend the GPS1, and release GPS1+ proper motion catalog to make up the limitations of {\it Gaia} DR2. Therefore, the GPS1+ will mainly focus on: (1) the sources ($19<G<20.7$\,mag), using the {\it Gaia} DR2 proper motions as priors to improve the proper motions combining PS1 and SDSS if their proper motions were measured in {\it Gaia} DR2; (2) the part of missing sources ($>361$ million) in {\it Gaia} DR2, their proper motions are calculated with the same procedure of GPS1; (3) the faint sources ($20.7 < G < 22.5$\,mag), using the same procedure as GPS1.

With the above motivations, we arrange the remainder of this paper as follows. In Section 2, we describe how to construct the GPS1+ catalog. In this section, we first summarize the four data sets in brief, and describe a Bayesian model to calculate proper motions for the sources which have {\it Gaia} proper motions. Section 3 then presents the results of GPS1+ proper motions, and demonstrate their performance in accuracy and precision. In Section 4, we briefly discuss the limitations of GPS1+ and summarize in Section 5. 

Throughout the paper, we adopt the Solar motion as $(U_\odot,V_\odot,W_\odot)=(9.58, 10.52, 7.01)$\,\kms\ \citep{tian2015}, and the IAU circular speed of the local standard of rest (LSR) as $v_0=220$\,\kms. Also, $\alpha*$ is used to denote the right ascension in the gnomonic projection coordinate system, for example, \mura\ $=$ $\mu_{\alpha}\cos(\delta)$, and $\Delta\alpha*=\Delta\alpha\cos(\delta)$, while ${\epsilon}$ denotes uncertainties, to avoid confusion with the symbol $\delta$ referring to a source's declination. We use $\Delta$ to denote the differences in quantities such as proper motion or position.


\section{The Construction of GPS1+}\label{sect:data}
\subsection{Data Set}\label{sect:data}
We still use the four basic imaging surveys, i.e., {\it Gaia}, PS1, SDSS, and 2MASS, to build the GPS1+ catalog. Unlike GPS1,  GPS1+ will be based on the {\it Gaia} DR2, but the other three astrometric datasets keep the same as those used in GPS1, i.e., the same data version and treatment. 

\begin{figure}[!t]
\centering
\includegraphics[scale=0.65]{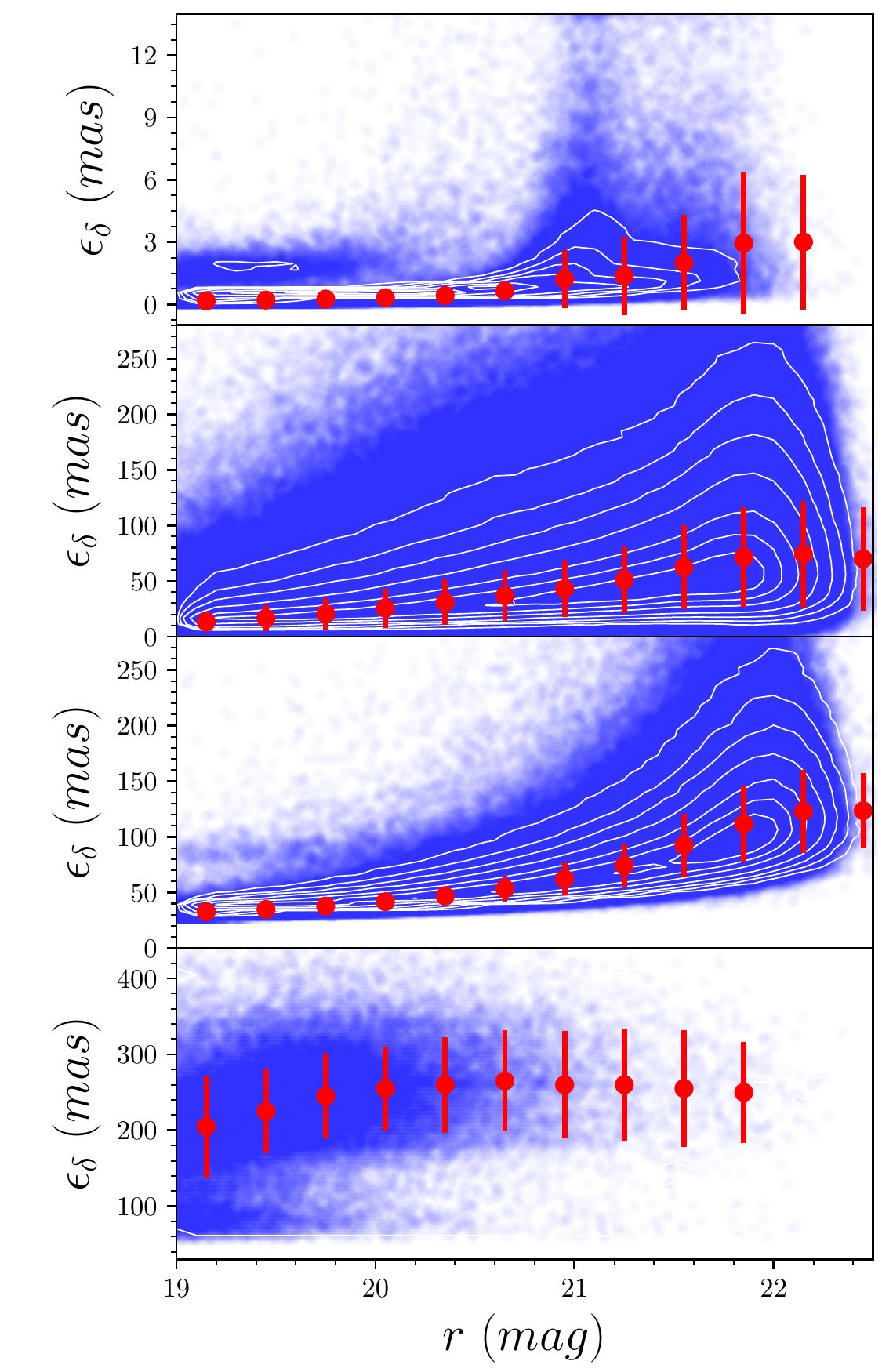}
\caption{Precision of the source position measurements along the $\delta$ direction for the various data sets used in the construction of the GPS1+ catalog, as a function of $r$-band magnitude. The red dots and bars indicate the average and root-mean-square ({\it rms}) of the position uncertainties in each magnitude bin. The average position uncertainties are 0.5, 50, 80, and 235\,mas for the entire samples in $19<r<22.5$\,mag from {\it Gaia} DR2, PS1, SDSS, and 2MASS, respectively. The contours indicate the normalized number density of sources with different levels of 0.02, 0.05, 0.1, 0.2, 0.4, 0.6, and 0.8 (the highest density is normalized to 1).}\label{fig:mr_raErr}
\end{figure}

{\it Gaia} DR2 consists of around 1.69 billion astrometric sources \citep{gaia2018}. All the sources have positions, and they are calibrated to the International Celestial Reference Frame (ICRF) at epoch J2015.5. The typical uncertainties in positions are the order of 0.7\,mas\ for sources at the faint end (i.e., G $=$ 20\,mag), as shown in the top panel of Figure \ref{fig:mr_raErr}. Therefore, {\it Gaia} DR2 is able to provide one precise observational position at epoch J2015.5. The epoch is different from J2015.0 in {\it Gaia} DR1.
 
\begin{figure}[!t]
\centering
\includegraphics[scale=0.5]{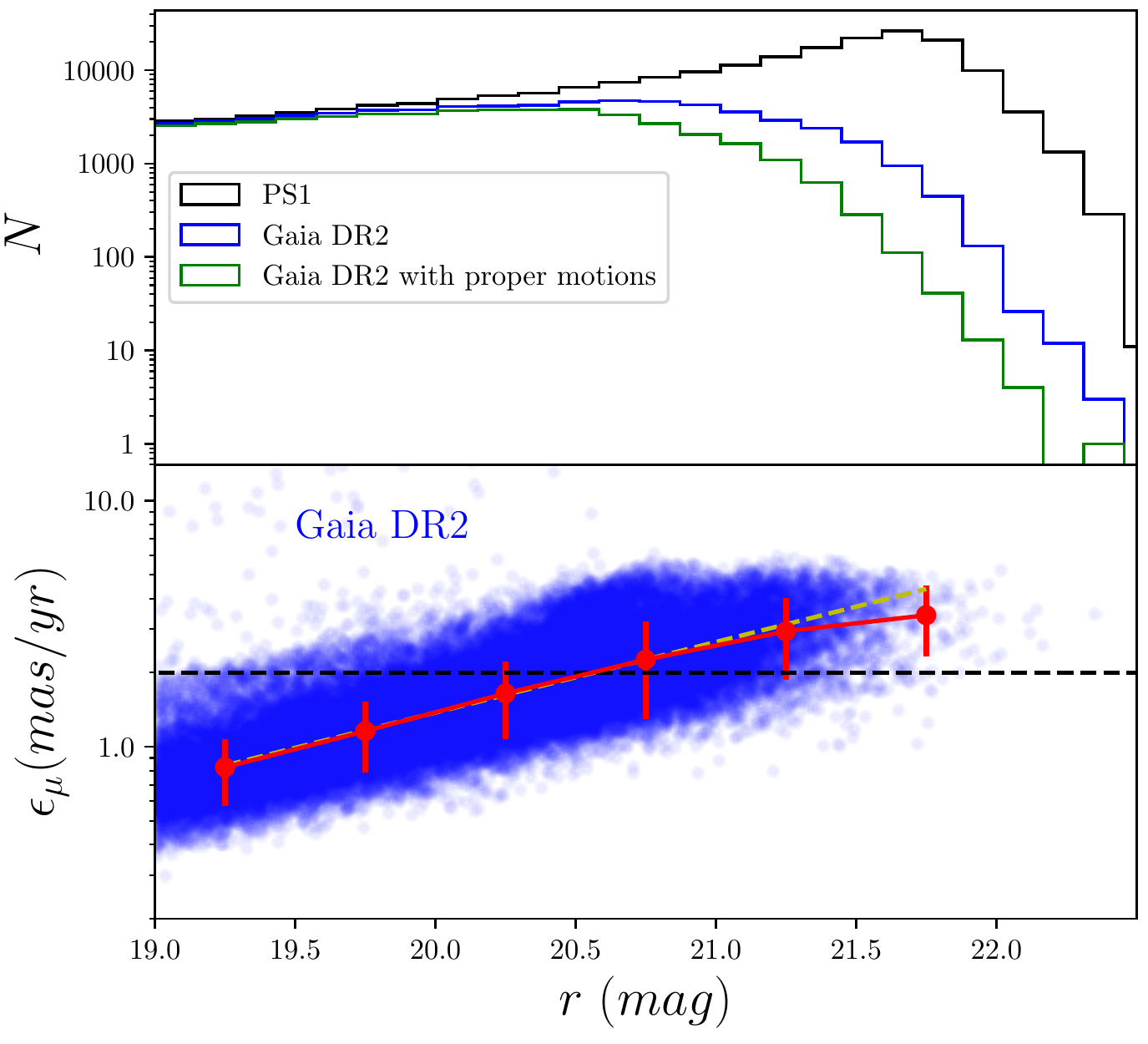}
\caption{Top: the histograms of the sources at the faint region ($r>19$\,mag) from {\it Gaia} DR2 (blue), PS1 (black), and {\it Gaia} DR2 in which sources have proper motions (green), in the same sky region. The sources with {\it Gaia} proper motions only take fraction of $\sim$40\% in a sample selected from PS1 in a random sky region. Bottom: the uncertainties distribution of {\it Gaia} DR2 proper motions at the faint region ($r>19$\,mag). The uncertainties increase with magnitudes as a function of a power law, i.e., {\bf the flux in r band to the 0.29 power} (the yellow dashed line). The average uncertainty of the proper motions is larger than 2 \masyr\ (marked with the black dashed line) for the sources nearby the {\it Gaia} limiting magnitude.}\label{fig:mr_gaia}
\end{figure}

About 1.33 billion sources have proper motions, but more than 361 million sources have no proper motions in {\it Gaia} DR2. The sources missing proper motions are mainly located at the faint region in {\it Gaia} DR2, as shown in the top panel of Figure \ref{fig:mr_gaia} by comparing the histograms between the entire sources (blue) and those without proper motions (green) in {\it Gaia} DR2. The bottom panel of Figure \ref{fig:mr_gaia} displays the scatter distribution of the uncertainties of {\it Gaia} DR2 proper motions at the faint region ($r>19$\,mag). The red points are the median uncertainties of proper motions in different magnitude bins. The median uncertainty is larger than 2.0 \masyr\ (marked with the black dashed line) for the sources close to the limiting magnitude. The proper motion precision of these sources will be significantly improved by combining the astrometry of PS1, SDSS and Gaia. This point will be demonstrated in Section \ref{sect:method}.


Pan-STARRS1 \citep[PS1;][]{Chambers2011} is a wide-field optical/near-IR survey telescope system, which has been conducting multi-epoch and multi-color observations over the entire sky visible from Hawaii (decl. $\gtrsim -30^\circ$) for many years. Its Processing Version 3 catalog \citep[PV3;][]{Chambers2016} contained around 65 detections for each source over a sky area of $\sim$30,000\ deg$^2$ with epochs throughout the 5.5 years from 2010 to 2014. 

As does in GPS1, we take season averages and positional uncertainties for the faint sources ($r_{PS1}>19.0$\,mag) in PS1. Each source is detected more than 10 times in an observing season. The typical single-epoch positional precision of faint sources is $\sim 50\,$ mas, as illustrated in the second panel of Figure \ref{fig:mr_raErr}. Furthermore, we apply the selection cuts used in GPS1 on the individual detections and the individual faint sources to remove the PS1 astrometry outliers. Finally, we obtain around 400 million faint objects with billions of detections.

The Sloan Digital Sky Survey (SDSS) began its regular operations in 2000 April \citep{York2000}. Its ninth data release (DR9) almost contains its all the photometric data \citep{Ahn2012}, which were imaged in the early epochs, e.g. 10-20 years ago. The long epoch baseline makes this data very valuable. The typical astrometric uncertainties for faint stars ($r>19.0$ mag) are around 80\,mas per coordinate \citep{Stoughton2002}, as shown in the third panel of Figure \ref{fig:mr_raErr}. 

Two Micron All Sky Survey \citep[2MASS;][]{Skrutskie2006} All-Sky Data Release identifies around 471 million point sources, and 1.6 million extended sources, covering virtually the entire celestial sphere between June 1997 and February 2001. Faint source extractions have the astrometric accuracy of the order 100\,mas, as shown in the bottom panel of Figure \ref{fig:mr_raErr}. Because of large positional uncertainties, 2MASS positions provide only a weak constraint for proper motion measurements.

We cross-matched the PS1 objects with {\it Gaia}, 2MASS and SDSS using a 1.5$\arcsec$ search radius. Therefore, the internal ID from PS1 is the key identifier to connect the four catalogs.

The black histogram in Figure \ref{fig:mr_gaia} indicates that there are more than 60\% PS1 sources beyond {\it Gaia}'s limiting magnitude. These sources ($>$70\%) without {\it Gaia} (including {\it Gaia} missing) proper motions will be our main interest in this study. 

\subsection{Derivation of Proper Motions}\label{sect:method}
Proper motions in GPS1+ are determined basically with the same procedure used in GPS1. The key difference just takes place on the sources which have {\it Gaia} proper motions. For these sources, we calculate two kinds of proper motions for each source: one is with the method of GPS1, i.e., by performing a linear least-squares fit; the other is fitted through a Bayesian model which uses {\it Gaia} proper motions as priors and combines the astrometry of {\it Gaia}, PS1, SDSS and 2MASS. It is worth to mention that both of the two fit methods do not involve Gaia parallax, so the derived proper motions will be away from the impact of parallax,  unlike {\it Gaia}'s proper motions which is correlated with the parallax.




With the procedure of GPS1 construction, we build a reference catalog by averaging repeatedly observed positions of PS1 galaxies in each tile (i.e., a sky area of a constant size of 10\degr\ by 10\degr), and calibrate the cataloged positions for each object in five (or six) PS1 epochs, one {\it Gaia} epoch, possibly one SDSS epoch and one 2MASS epoch onto the same reference frame. All the steps have been minutely summarized in Section 3 of T17. For each source, we calculate its proper motion by performing a linear least squares fit based on a simple $\chi^2$ which is described in Equation (2) of T17. But for a source which has a {\it Gaia} proper motion, we re-measure another new proper motion by building a Bayesian model. We start with a likelihood
\begin{equation}\label{eq:bayesian}
L=p(\{t_i,y_i\} | {\mu}, {b})=\prod_{i}^{N}\left\{\frac{1}{\sqrt{\epsilon_{i}^2}}\exp\Bigl[-\frac{[\hat{y}_{i}^{o} - y_{i}^{model}(t_i)]^2}{2\epsilon_{i}^2}\Bigr]\right\},
\end{equation}
where $\hat{y}_i^o$ is the observed position of a star with a positional uncertainty $\epsilon_{i}$ at epoch $i$. The positional uncertainty $\epsilon_{i}$ consists of two parts: one part is the individual position precision, illustrated in Figure \ref{fig:mr_raErr}; and the other part is the uncertainty from the offset calibration discussed in Section 3.2 of T17. $y_i^{model}(t_i)$ is the predicted position by a linear model at the given time $t_i$, i.e., $y_i^{model}(t_i) = {\mu}t_i + {b}$, $N$ is the number of epochs in different surveys.
The position $\hat{y}_i^o$ has been calibrated by
\begin{equation}\label{eq:cal_offset}
\hat{y}_i^o = y_i^o - \Delta_{i}(\alpha, \delta) - \Delta_{i}(\delta, m),
\end{equation}
where $y_i^o$ is the original cataloged position of a star at epoch $i$, $\Delta_{i}(\alpha, \delta)$ is the direction dependent offset described in Section 3.2.1 of T17, and $\Delta_{i}(\delta, m)$ is the magnitude and declination dependent offset described in Section 3.2.2 of T17.  

According to Bayes theorem, the posterior probability can be easily expressed as
\begin{equation}\label{eq:posterior}
p(\{{\mu}, {b} | t_i,y_i\})=p(\{t_i,y_i\} | {\mu}, {b}) p(\mu) p(b) ,
\end{equation}
we assume the prior probability of $\mu$ obeys a Gaussian distribution with $\bar\mu = \mu_{Gaia}$ and $\sigma=\epsilon_{Gaia}$, where $\mu_{Gaia}$ and $\epsilon_{Gaia}$ are the proper motion and uncertainty values of a source provided by {\it Gaia} DR2. We assume a flat prior probability of $b$, i.e., $p(b) = 1$.

\begin{figure*}[!t]
\centering
\includegraphics[width=0.48\textwidth, trim=0.0cm 0.0cm 0.0cm 0.0cm, clip]{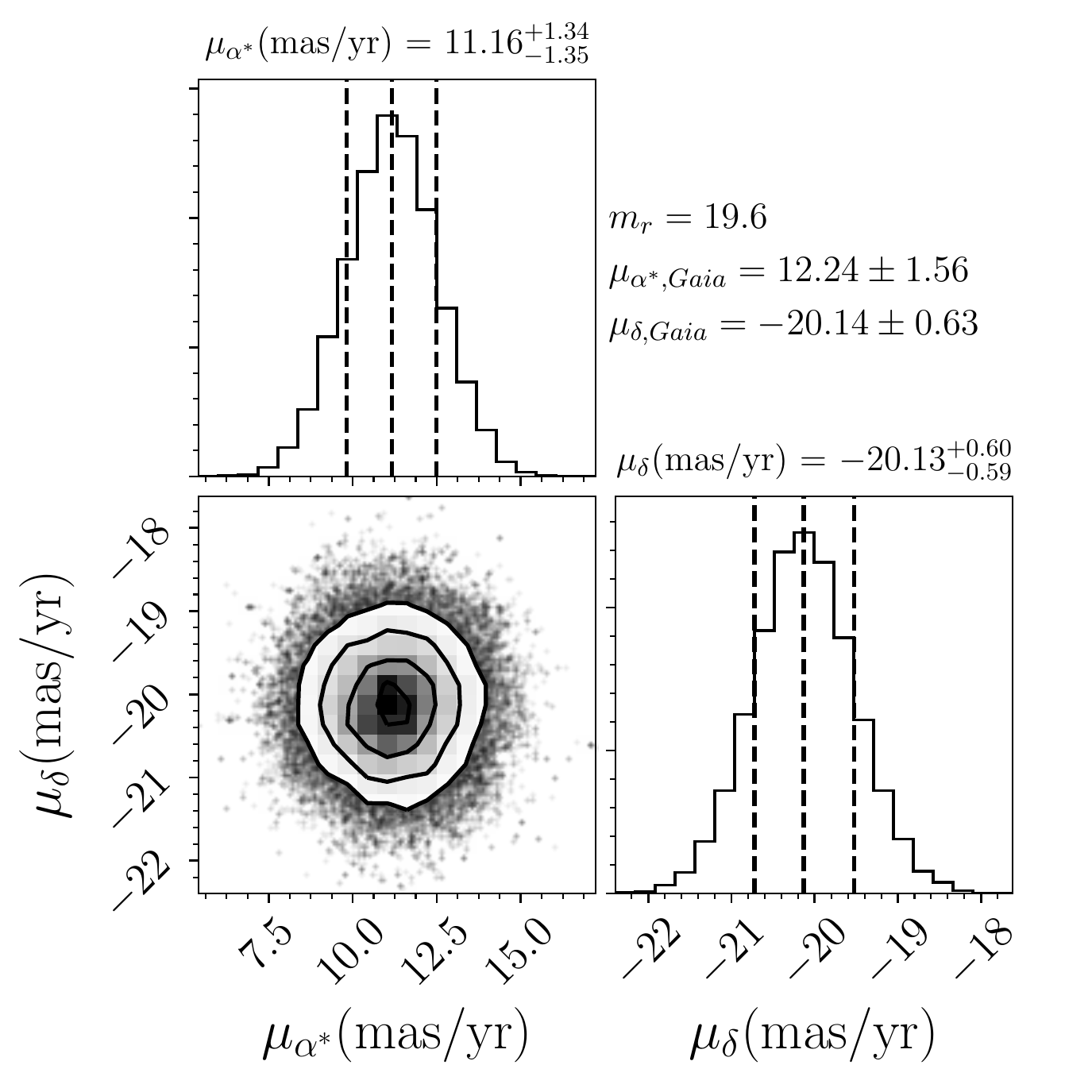}
\includegraphics[width=0.48\textwidth, trim=0.0cm 0.0cm 0.0cm 0.0cm, clip]{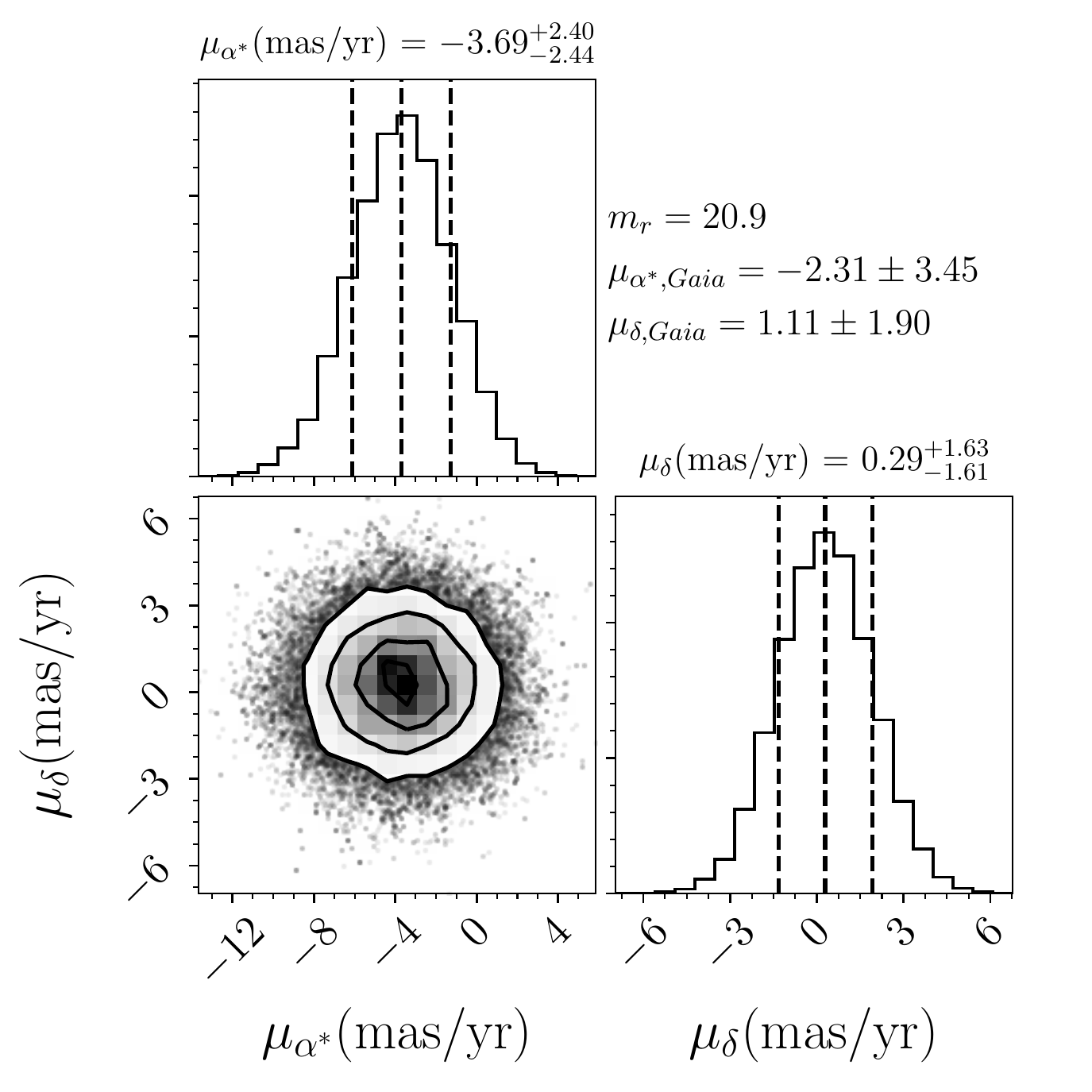}
\caption{Illustration of the proper motion contours and marginalized probability distributions for two sources with different magnitudes, from the MCMC sampling. The left panel shows the case with a magnitude of $r=19.6$\,mag. In this case, the source is not too faint, the {\it Gaia} detector works well. Therefore, the proper motions ($12.24\pm1.56$\masyr\ and  $-20.14\pm0.63$\masyr) are well measured in {\it Gaia} DR2. The precision of new proper motions ($11.16\pm1.34$\masyr\ and  $-20.13\pm0.60$\masyr) can not be improved significantly (only by $\Delta\epsilon_{\mu}\sim0.2$\masyr) via combining the astrometry from SDSS, PS1, and Gaia. The right panel displays the case with a magnitude of $r=20.9$\,mag, which is close to the {\it Gaia} limiting magnitude. In this case, the precision of proper motion can be improved significantly (by $\Delta\epsilon_{\mu}\sim1.0$\masyr) with the combination of the multi-surveys.
 }
   \label{fig:mcmc}
\end{figure*}

We use \texttt{emcee} \citep{forman2013} to sample the posterior distribution (Equation \ref{eq:posterior}) and estimate proper motions in the two directions, i.e., \mura\ and \mudec, respectively. In practice, we could use the joint posterior probability to constrain \mura\ and \mudec, simultaneously. The intercept $b$ is also a free parameter in the MCMC sampling, but its value is not important for this study. Figure \ref{fig:mcmc} illustrates two examples of proper motion contours and marginalized probability distributions of two sources with different magnitudes. The {\it Gaia} detector takes on different performances for sources with distinct brightness. For instance, {\it Gaia} is able to measure a good position for a source with $r=19.6$\,mag. Thus, the combination of the multi-surveys can not significantly improve the precision of the {\it Gaia} proper motion (only by $\Delta\epsilon_{\mu}\sim0.2$\masyr, see the left panel of Figure \ref{fig:mcmc}). However, for a source with $r=20.9$\,mag, which is close to the {\it Gaia} limiting magnitude, the combination of PS1, SDSS and {\it Gaia} can improve the precision of the {\it Gaia} proper motion by $\Delta\epsilon_{\mu}\sim1.0$\masyr (see the right panel of Figure \ref{fig:mcmc}).

\section{Results and performance}\label{sect:result}

Using the approach described in Section \ref{sect:method} and the method used in GPS1 (see Section 3 of T17), we determine proper motions for around 400 million sources, down to a magnitude of $\sim 22.5$\, in the r-band. Among these sources, about 40\% sources are re-measured new proper motions with the Bayesian method described in Section \ref{sect:method}. The proper motions of the remaining objects are obtained with the previous method used in GPS1. The catalog draws on PS1 SeasonAVG and {\it Gaia} DR2 as the primary data, together with the best available combinations of other surveys. The final catalog uses the robust fit (where all the data points are fitted regardless of outliers), cross-validation fit (where outliers are removed while fitting), and MCMC fit (with which proper motions from {\it Gaia} DR2 are used as priors while sampling if the proper motions exist in {\it Gaia} DR2). For reference, we also include the proper motions of {\it Gaia} DR2 if they exist. Table \ref{tab:gps1} lists the main columns contained in the catalog. In the following sub-sections, we discuss the precision and accuracy of proper motions in the different cases.

\begin{table*}
 \begin{threeparttable}
\caption{The columns of GPS1+ catalog}.\label{tab:gps1}
\centering
\begin{tabular}{l|l|l|l}
\hline
\hline
&Column&Unit&description\\
\hline
1&obj\_id\footnotemark&-&The unique \HJT{but internal} object\_id in PS1\\
2&ra&degree&R.A. at J2015.0 from {\it Gaia} DR2\\
3&dec&degree&Decl. at J2015.0 from {\it Gaia} DR2\\
4&e\_ra&mas&Positional uncertainty in right ascension at J2015.0 from {\it Gaia} DR2\\
5&e\_dec&mas&Positional uncertainty in declination at J2015.0 from {\it Gaia} DR2\\
6&ra\_ps1&degree&Average right ascension at J2010 from PS1 PV3\\
7&dec\_ps1&degree&Average declination at J2010 from PS1 PV3\\
8&pmra&\masyr&Proper motion with robust fit in $\alpha\cos\delta$\\
9&pmde&\masyr&Proper motion with robust fit in $\delta$\\
10&e\_pmra&\masyr&Error of the proper motion with robust fit in $\alpha\cos\delta$\\
11&e\_pmde&\masyr&Error of the proper motion with robust fit in $\delta$\\
12&chi2pmra&-&$\chi_{\nu}^2$ from the robust proper motion fit in $\alpha\cos\delta$\\
13&chi2pmde&-&$\chi_{\nu}^2$ from the robust proper motion fit in $\delta$\\
14&pmra\_x&\masyr&Proper motion with cross-validated fit in $\alpha\cos\delta$\\
15&pmde\_x&\masyr&Proper motion with cross-validated fit in $\delta$\\
16&e\_pmra\_x&\masyr&Error of the proper motion with cross-validated fit in $\alpha\cos\delta$\\
17&e\_pmde\_x&\masyr&Error of the proper motion with cross-validated fit in $\delta$\\
18&pmra\_mcmc&\masyr&Proper motion with MCMC sampling fit in $\alpha\cos\delta$\\
19&pmde\_mcmc&\masyr&Proper motion with MCMC sampling fit in $\delta$\\
20&e\_pmra\_mcmc&\masyr&Error of the proper motion with MCMC sampling fit in $\alpha\cos\delta$\\
21&e\_pmde\_mcmc&\masyr&Error of the proper motion with MCMC sampling fit in $\delta$\\
22&pmra\_gaia&\masyr&Proper motion from {\it Gaia} DR2 in $\alpha\cos\delta$\\
23&pmde\_gaia&\masyr&Proper motion from {\it Gaia} DR2 in $\delta$\\
24&e\_pmra\_gaia&\masyr&Error of the proper motion from {\it Gaia} DR2 in $\alpha\cos\delta$\\
25&e\_mude\_gaia&\masyr&Error of the proper motion from {\it Gaia} DR2 in $\delta$\\
26&n\_obsps1&-&The number of SeasonAVG observations used in the proper motion fit \\
27&n\_obs&-&The number of all the observations used in the robust proper motion fit\\
28&flag\footnotemark&-&An integer number used to flag the different data combination in the proper motion fit. \\
29&magg&mag&g-band magnitude from PS1 \\
30&magr&mag&r-band magnitude from PS1\\
31&magi&mag&i-band magnitude from PS1\\
32&magz&mag&z-band magnitude from PS1\\
33&magy&mag&y-band magnitude from PS1\\
34&e\_magg&mag&Error in g-band magnitude from PS1\\
35&e\_magr&mag&Error in r-band magnitude from PS1\\
36&e\_magi&mag&Error in i-band magnitude from PS1\\
37&e\_magz&mag&Error in z-band magnitude from PS1\\
38&e\_magy&mag&Error in y-band magnitude from PS1\\
39&maggaia&mag&G-band magnitude from {\it Gaia} \\
40&e\_maggaia&mag&Error in G-band magnitude from Gaia\\
\hline
\hline
\end{tabular}
 \begin{tablenotes}
  \item [a] \HJT{Here obj\_id is an internal PS1 ID, which is different from the public ID released in PS1 catalog.}
 \item [b] \HJT{In order to label the different survey combinations for proper motion fit, we assign PS1, 2MASS, SDSS, and {\it Gaia} with different integer identifiers, i.e. 0, 5, 10, and 20, respectively, and define a $flag$ with the sum of identifiers of surveys combined.}
  \end{tablenotes}
 \end{threeparttable}
\end{table*}

\subsection{Proper Motion Uncertainties in the Different Data Set Combinations}

The \HJT{footprint overlap} among {\it Gaia}, PS1, SDSS and 2MASS surveys introduces some complexity: $\sim$ 14.5\% stars are covered by {\it Gaia}, PS1, and SDSS, $\sim$ 43.6\% by PS1 and {\it Gaia}, but not SDSS, $\sim$ 33.9\% stars are only observed by PS1, and the remaining 8\% by PS1 and SDSS, but not Gaia. Therefore, it is necessary to investigate how the final proper motions are affected by combining the different data sets.

\begin{table*}
 \begin{threeparttable}[b]
\caption{The formal fitting uncertainties of the proper motions in the different data combinations}.\label{tab:dif_comb}
\centering
\begin{tabular}{c|l|c|c|c|c|c|c}
\hline
\hline
ID&Mode& \multicolumn{2}{c|}{$19<m_r<21$} &\multicolumn{2}{c|}{$21<m_r<22.5$}&\multicolumn{2}{c}{MCMC Sampling\tnote{a}} \\
\hline
&&\mmura &\mmudec &\mmura &\mmudec &\mmura &\mmudec \\
\hline
\multicolumn{2}{c|}{}&\multicolumn{6}{c}{\masyr}\\
\hline
1&GPS (Gaia+PS1+SDSS+2MASS)&2.10$\pm$0.64&1.97$\pm$0.60&3.18$\pm$0.94&2.93$\pm$0.82&0.96$\pm$0.50&0.85$\pm$0.47\\
2&GP  (Gaia+PS1+2MASS)&3.71$\pm$2.07&2.98$\pm$1.35&5.10$\pm$2.34&4.24$\pm$1.68&1.03$\pm$0.61&0.95$\pm$0.58\\
3&PD  (PS1+SDSS+2MASS)&5.12$\pm$2.33&4.85$\pm$2.22&7.56$\pm$3.26&7.16$\pm$3.05&-&-\\
4&PS1 ({\sl only} PS1)&15.20$\pm$9.93&12.75$\pm$6.67&21.58$\pm$12.67&18.25$\pm$9.72&-&-\\
\hline
\hline
\end{tabular}
 \begin{tablenotes}
 \item [a] The uncertainties in this column are estimated from the sources with $19.0<r<22.5$\,mag. In this mode, proper motions are measured with {\it Gaia} proper motions as priors during MCMC sampling. Therefore, there are no values in the PD and {\sl only} PS1 modes.
  \end{tablenotes}
 \end{threeparttable}
\end{table*}

Like GPS1, we investigate how the uncertainties in proper motion differ among the following four combinations of data sets:  {\it Gaia} + PS1 + SDSS + 2MASS (GPS), {\it Gaia} + PS1 + 2MASS (GP), PS1 + SDSS + 2MASS (PD), and 
{\sl only} PS1 (PS1). \HJT{For the catalog table, different} surveys are assigned different integer \HJT{identifiers: 0, 5, 10, and 20 for PS1, 2MASS, SDSS, and {\it Gaia}, respectively. This defines a {\sl flag} for different survey combinations entering a fit}, represented as the sum of \HJT{the individual survey identifiers}. The primary observations are those from PS1, so the positions for each star must include the PS1 detections when fitting for proper motion.

\begin{figure*}[!t]
\centering
\includegraphics[width=0.504\textwidth, trim=0.0cm 0.0cm 0.1cm 0.0cm, clip]{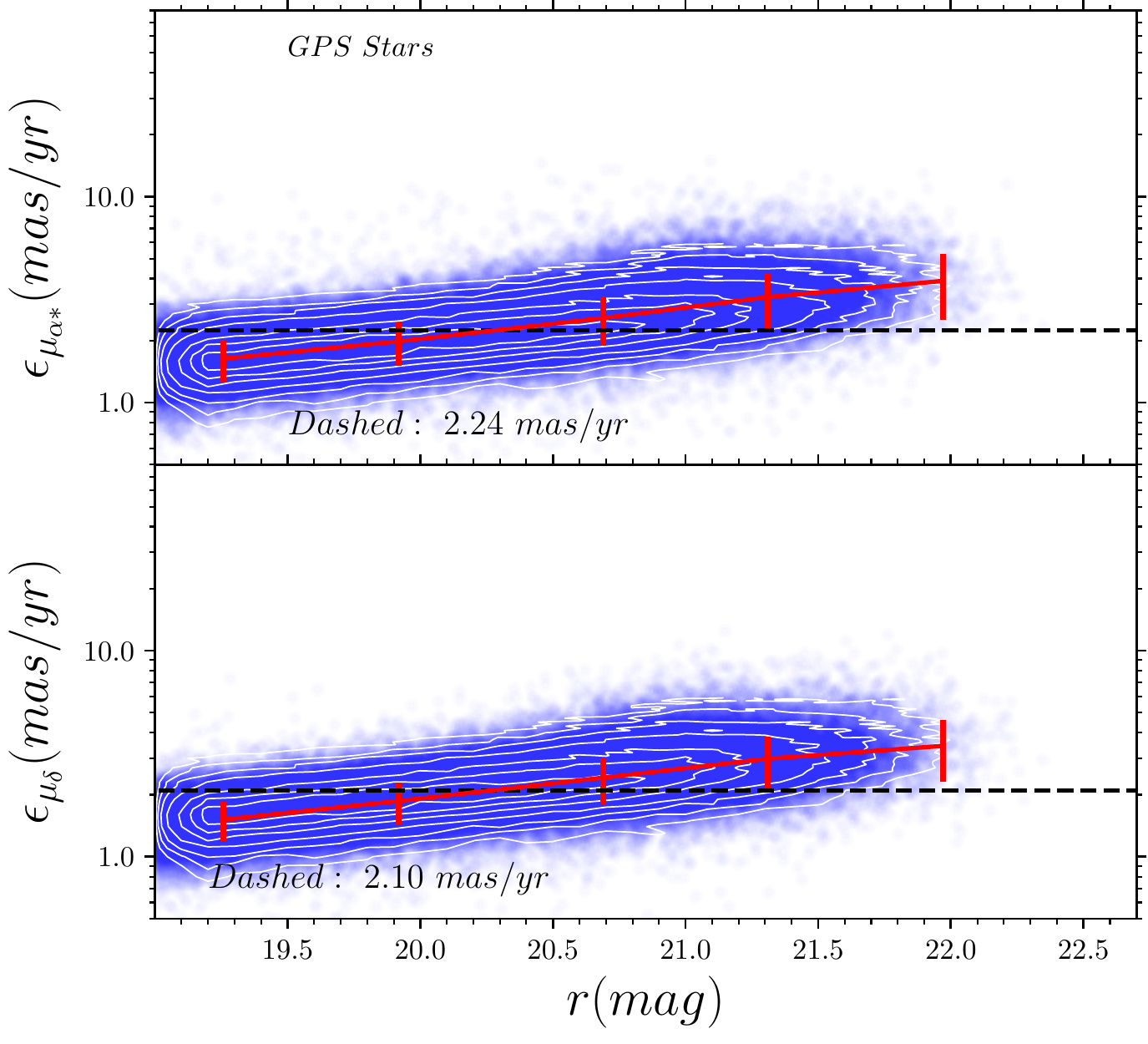} 
\includegraphics[width=0.44\textwidth, trim=1.85cm 0.0cm 0.0cm 0.0cm, clip]{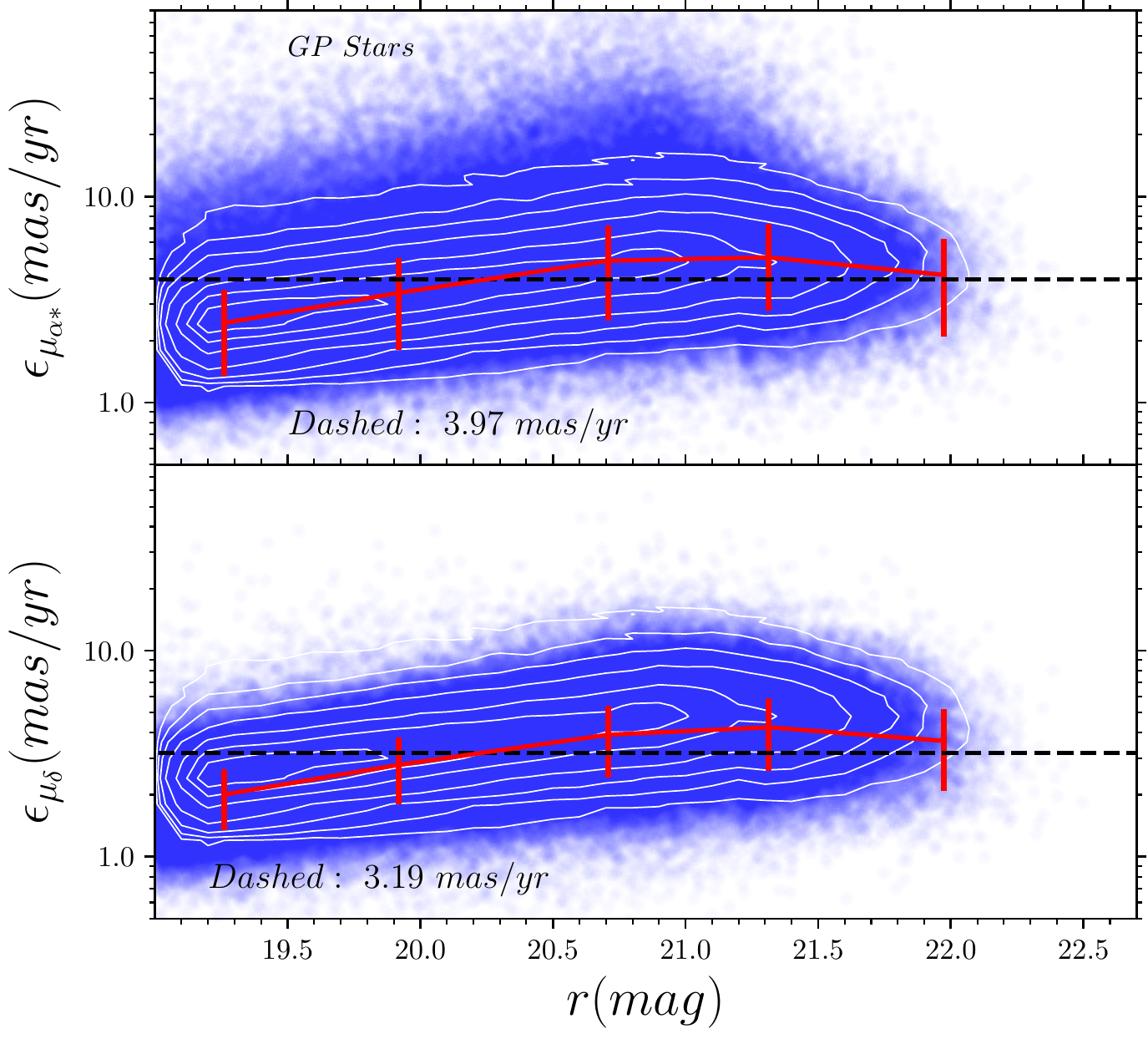} 
\includegraphics[width=0.504\textwidth, trim=0.0cm 0.0cm 0.1cm 0.0cm, clip]{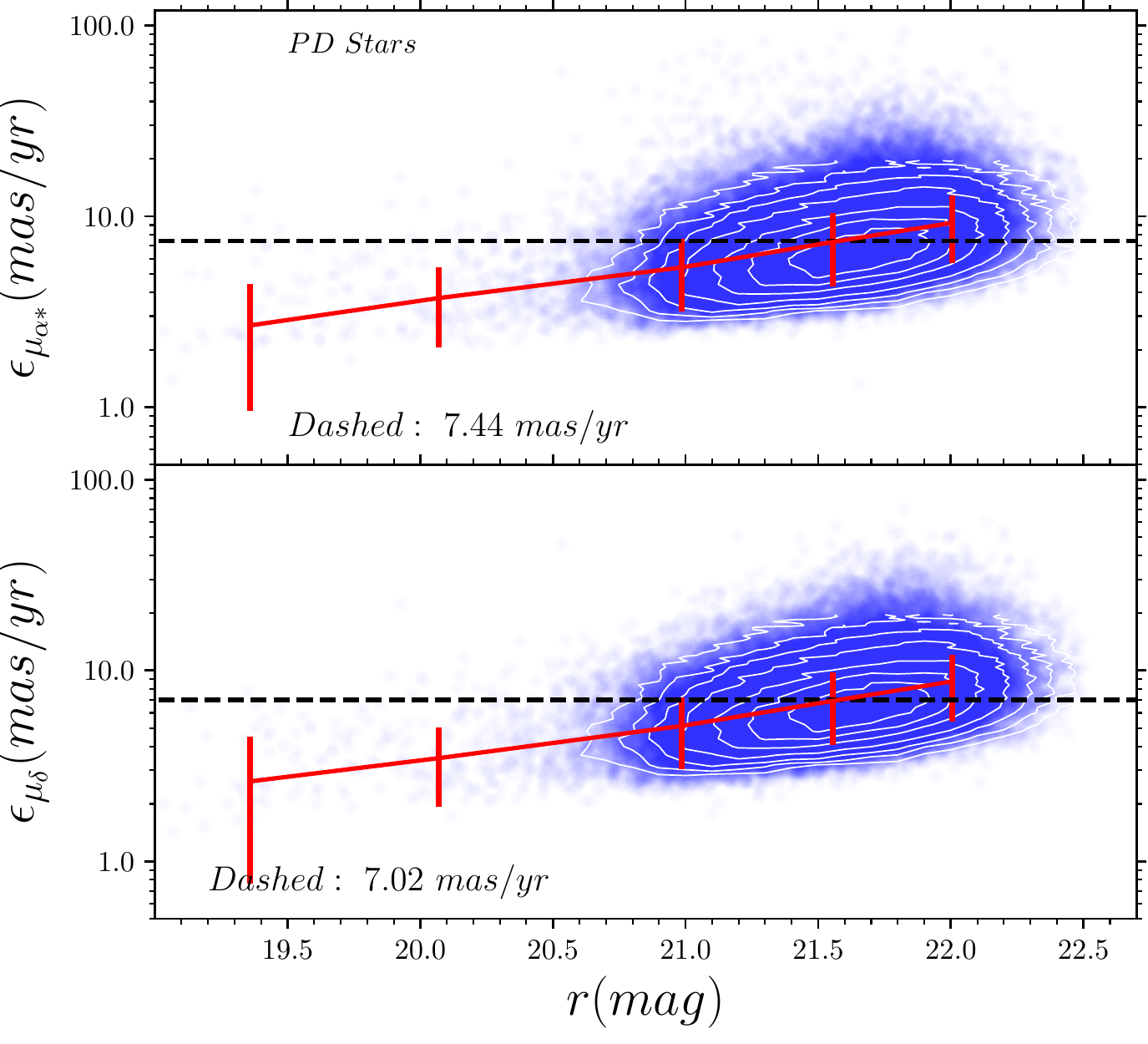} 
\includegraphics[width=0.44\textwidth, trim=1.85cm 0.0cm 0.0cm 0.0cm, clip]{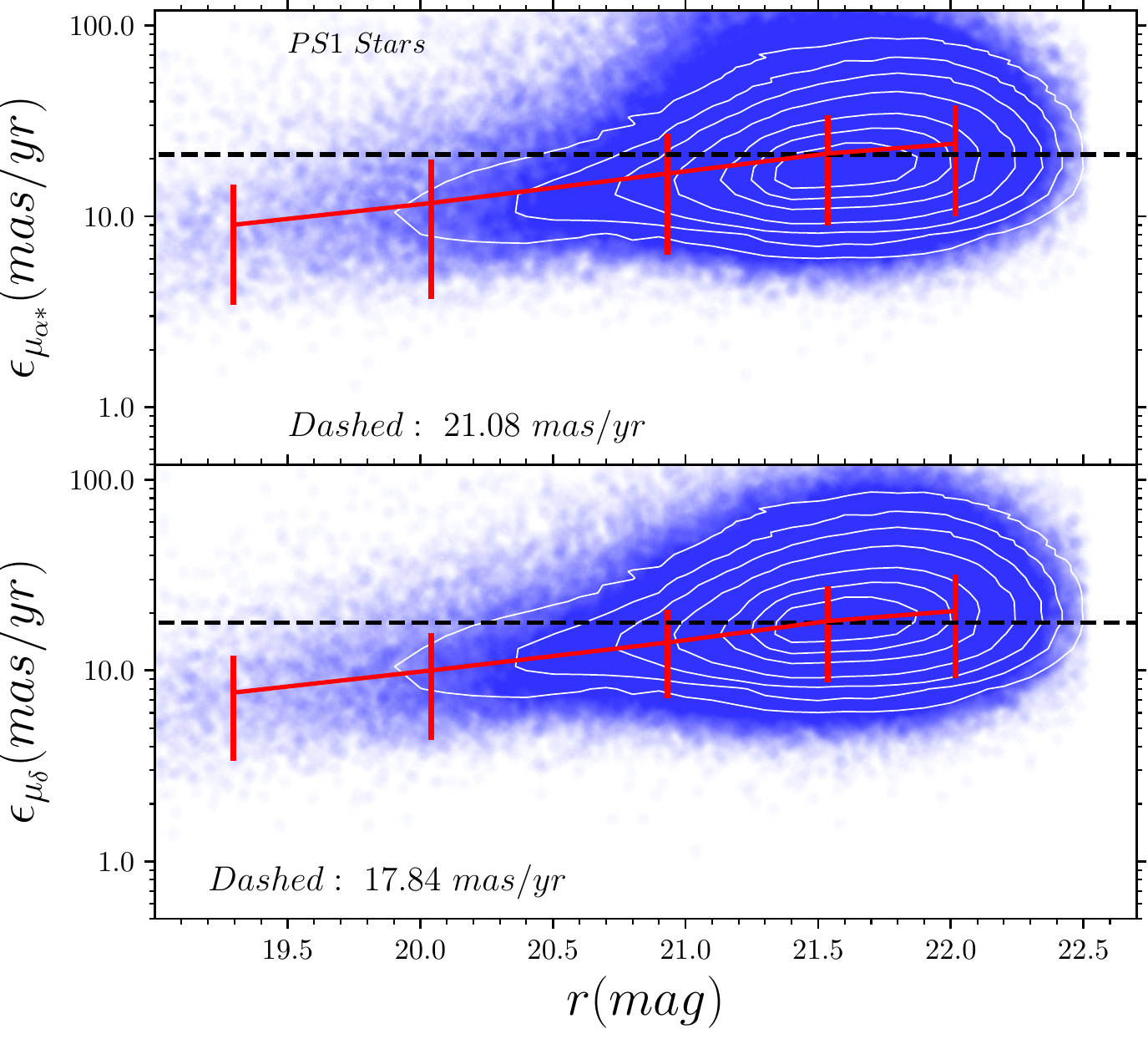} 
\caption{Proper motion precision for the four different combinations of data sets (top-left: GPS, top-right: GP, bottom-left: PD, and bottom-right: ONLY PS1). In the four panels, the red curves and bars are the median uncertainties and {\it rms} of proper motions within different magnitude bins, and the black dashed lines mark the typical average uncertainties in the magnitude range  $19<r<22.5$\,mag. The blue scatter points represent one million sources randomly selected from the sky. All the uncertainties are logarithmic in every y-axis.
The typical average uncertainties for the four combination modes are \dmura\ $\sim$2.24 \masyr, \dmudec\ $\sim$2.10 \masyr\ for the GPS mode, \dmura\ $\sim$3.98 \masyr, \dmudec\ $\sim$3.19 \masyr\ for the GP mode, \dmura\ $\sim$7.45 \masyr, \dmudec\ $\sim$7.05 \masyr\ for the PD mode, and \dmura\ $\sim$21.03 \masyr, \dmudec\ $\sim$17.75 \masyr\ for the ONLY PS1 mode, respectively. The contours indicate the normalized number density of sources with different levels of 0.02, 0.05, 0.1, 0.2, 0.4, 0.6, and 0.8 (the highest density is normalized to 1).
}
   \label{fig:sigmu_mr_GPS1}
\end{figure*}

Figure \ref{fig:sigmu_mr_GPS1} summarizes the distribution of proper motion uncertainties for the four different combinations. The figure is drawn with one million sources randomly selected from the GPS1+ catalog. In the four panels, the blue points correspond to the stars in different combinations and the red curves are the median uncertainties in proper motions within different magnitude bins. \HJT{The average uncertainties in magnitude bins are listed in Table \ref{tab:dif_comb}, with the mean ($19<m_r<22.5$) marked by black lines. In the GPS mode, the average uncertainties are \dmura\ $\sim$2.24 \masyr\ and \dmudec\ $\sim$2.10 \masyr. This is better than the GP mode (\mura\ $\sim$3.98 \masyr and \mudec\ $\sim$3.19 \masyr). SDSS positions improve the precision by $\sim1.5$\masyr\ for both the \dmura\ and \dmudec. Without {\it Gaia} positions (PD mode), the typical uncertainties become \dmura\ $\sim$7.45 \masyr and \dmudec\ $\sim$7.05 \masyr. {\it Gaia} positions improve the precision by $\sim4.3$\masyr\ for both \dmura\ and \dmudec. For PS1 data alone, the mean uncertainties become \dmura\ $\sim$21.03 \masyr and \dmudec\ $\sim$17.75 \masyr. The precision improvement is dominated by {\it Gaia} and SDSS.}  

\begin{figure*}[!t]
\centering
\includegraphics[width=0.45\textwidth, trim=0.0cm 0.0cm 0.0cm 0.0cm, clip]{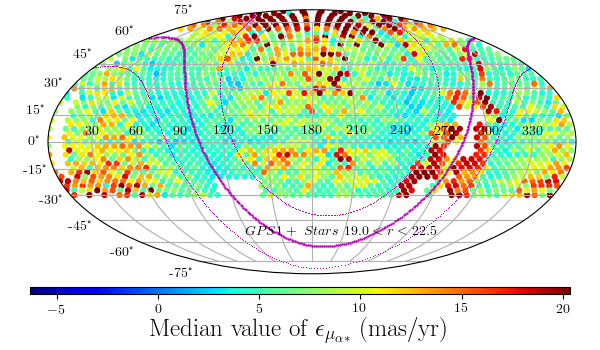}
\includegraphics[width=0.45\textwidth, trim=0.0cm 0.0cm 0.0cm 0.0cm, clip]{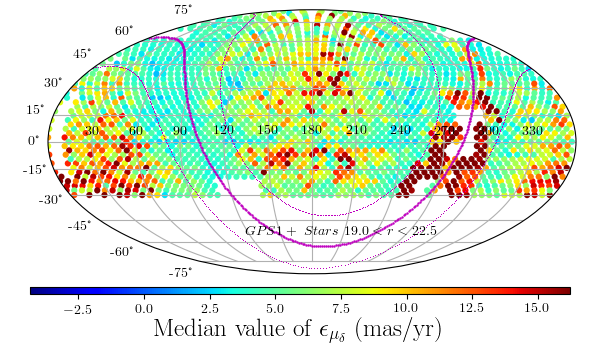}
\caption{The distribution of proper motion uncertainties for stars with $19<r<22.5$\,mag; this is illustrated with an equatorial Mollweide projection of the entire 3$\pi$ sky region. 
The pink solid ($b=0^{\circ}$) and two dotted lines ($b=\pm20^{\circ}$) mark the location of the Galactic plane in the equatorial coordinate system, \HJT{where sources are crowded and the effects of dust extinction are manifest \citep{tian2014}}. 
To highlight the structures in the maps, the color bar is scaled in $\pm3\sigma$ around the entire median value for each map. 
}\label{fig:uncertanties_star}
\end{figure*}

Figure \ref{fig:uncertanties_star} illustrates the distribution of uncertainties of these stars as Mollweide projection maps of the entire 3$\pi$ region of the sky in equatorial coordinate system, containing one million stars randomly selected. The median uncertainty in each pixel is calculated \HJT{from hundreds of stars}.
The median values of the uncertainties are $\sim8.0$\masyr\ for \mura\ (the left panel) and $\sim7.2$\masyr\ for \mudec\ (the right panel), as shown in the maps. The uncertainties at high and low declinations are larger, \HJT{as SDSS data are missing}. The small uncertainties in the north Galactic cap are driven by the SDSS observations taken ten or fifteen years ago. 

For sources with $19<r<21$\,mag, the GPS1+ catalog is at its best. In this magnitude bin, $\sim$92\% sources have {\it Gaia} positions, and $\sim$79.5\% sources have {\it Gaia} proper motions. Therefore, the proper motions in this bin are dominated by the values obtaining with MCMC fitting. During the fitting, we use {\it Gaia} proper motions as priors, and combine all the astrometry from {\it Gaia}, PS1, SDSS and 2MASS. The final uncertainties are better than 1.0 \masyr\ for both \dmura\ and \dmudec\ on average in this bin.

For the fainter sources with $r>21.0$\,mag, the positional uncertainties steeply increase with the magnitude. The magnitudes of these sources are beyond {\it Gaia} or close to PS1 and SDSS limiting magnitudes, so the precision of the obtained proper motions will be worse towards the faint end. As the values in Table \ref{tab:dif_comb} show, both SDSS and {\it Gaia} can improve the precision of the proper motions in the PS1 mode by $\sim10$\masyr\ individually, and by $\sim15$\masyr\ together. Therefore, {\it Gaia} and SDSS are comparably important for reducing uncertainties \HJT{for the faint stars}.

We checked the quality of the proper motion fits via the distribution of reduced $\chi^2$ for a random subset of stars. The median values for both \mura\ and \mudec\ are smaller than 1, implying that most fits are good.

According to the performance, the around 66\% sources in the GPS, GP, and PD modes are defined as the primary sources, which have a good precision with an average value of 2.0-5.0 \masyr; while the remaining $\sim$34\% sources only have PS1 astrometry, which are defined as the secondary sources with an average precision of worse than 15.0 \masyr. The bad precision makes the secondary sources probably have no good applications.

\subsection{Proper Motion Validation with QSOs}\label{sec:val}
To validate the derived proper motions, we cross-match the GPS1+ catalog with the QSO candidates from \citet{nina2016}, and randomly select 58,000 QSOs with high probability in the entire PS1 3$\pi$ sky region. 

\begin{figure}[!t]
\centering
\includegraphics[width=0.48\textwidth, trim=0.0cm 0.0cm 0.0cm 0.0cm, clip]{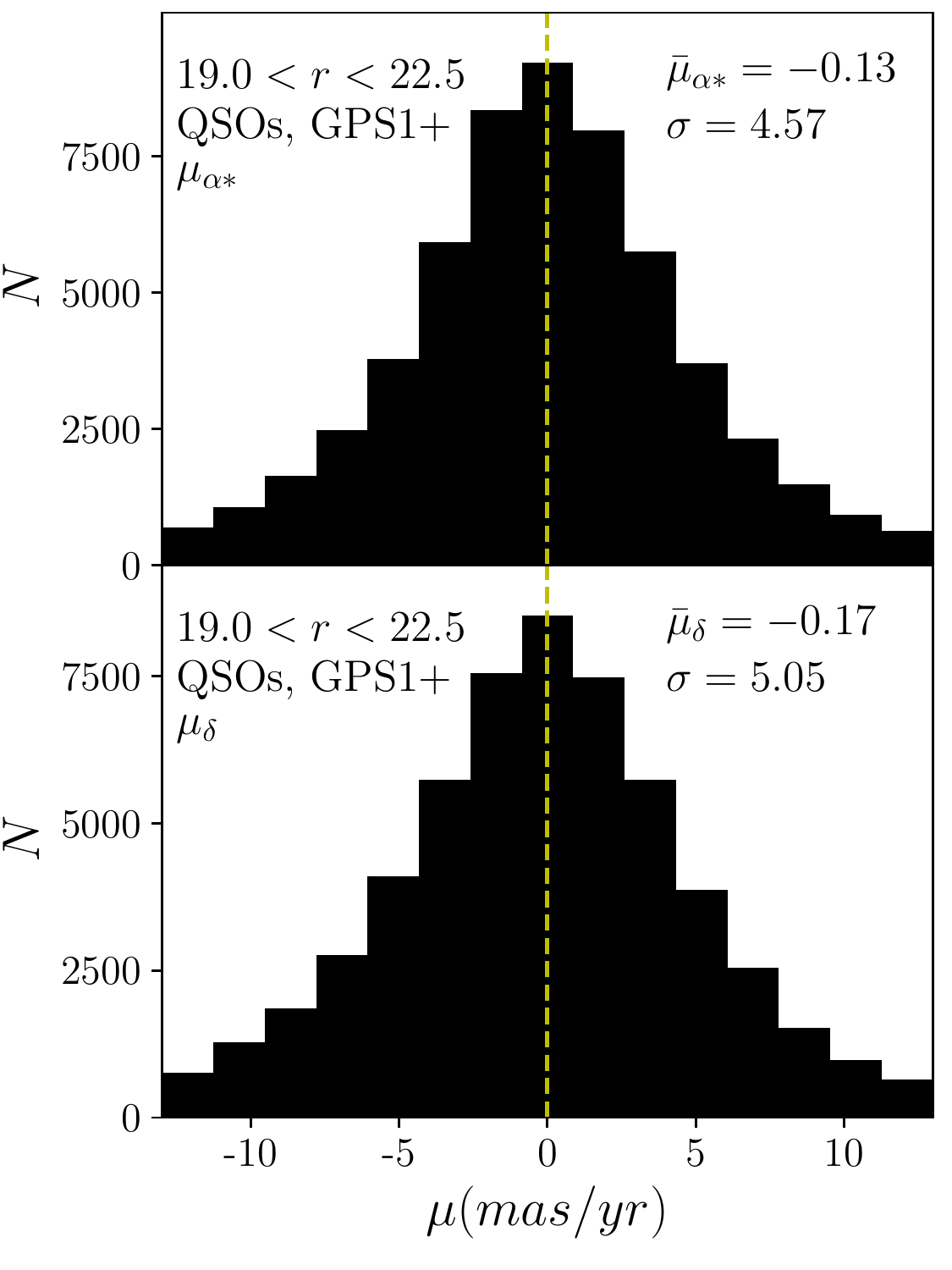}
\caption{Validation of GPS1+ proper motions with QSOs. The light dashed lines denote zero \masyr. 
}
   \label{fig:vali_qsos}
\end{figure}

Figure \ref{fig:vali_qsos} displays the histograms of the \mura\ (the top panel) and \mudec\ (the bottom panel) for the QSOs. The median values of the \mura\ and \mudec\ are -0.13 \masyr\ and -0.17 \masyr, and the dispersions are 4.57 \masyr\ and 5.05 \masyr, respectively. The median values suggest that the accuracies of GPS1+ proper motions are better than 0.2 \masyr\ on average for both \mura\ and \mudec. The dispersion values roughly reflect the {\it rms} of GPS1+ proper motions. Note that the apparent proper motions of QSOs suffer from the impact of differential chromatic refraction (DCR), especially in $\delta$. At high declinations, the $\delta$ proper motions are biased by up to 2\masyr. At low declinations, the $\delta$ proper motions are under-estimated by $\sim$ 2.0 \masyr. This definitely makes the dispersion values of the QSO proper motions become larger than the true values.

\subsection{Comparison with {\it Gaia} Proper Motions }\label{sect:val_pal5}
{\it Gaia} DR2 provides us enough proper motions with good measurements for stars in the entire sky. In GPS1+, we re-calculate the proper motions for the sources with {\it Gaia} DR2 proper motions in two methods: (1) the GPS1 method, in which proper motions are obtained by combining the astrometry of {\it Gaia}, PS1, SDSS, and 2MASS, regardless of {\it Gaia} DR2 proper motions; (2) the MCMC method, in which proper motions are obtained by combining the astrometry of {\it Gaia}, PS1, SDSS, and 2MASS, and using {\it Gaia} DR2 proper motions as priors during MCMC sampling. For the comparison, we randomly select about half a million of stars which have three kinds of proper motions, simultaneously.

\begin{figure*}[!t]
\centering
\includegraphics[width=0.45\textwidth, trim=0.0cm 0.0cm 0.0cm 0.0cm, clip]{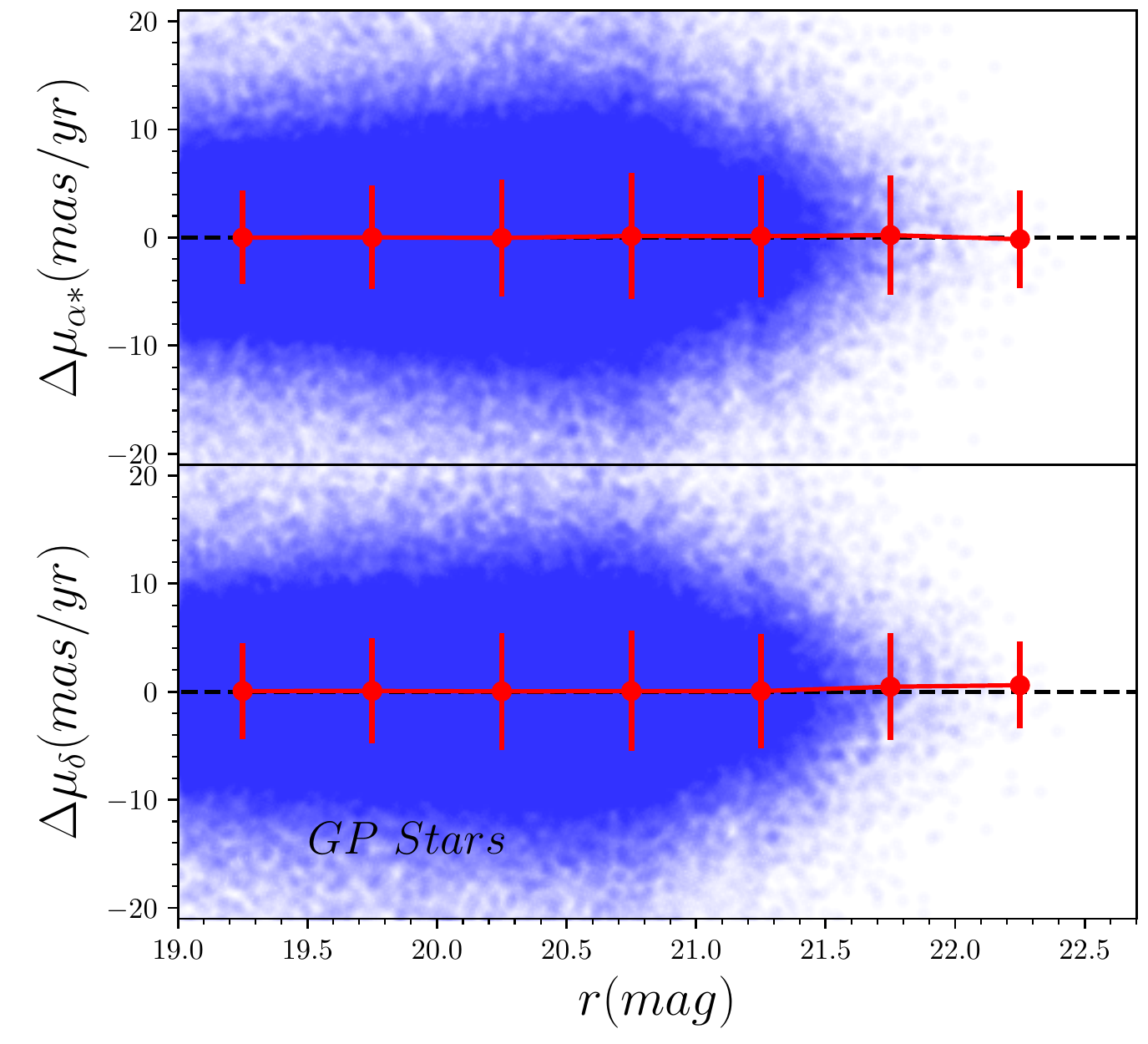}
\includegraphics[width=0.45\textwidth, trim=0.0cm 0.0cm 0.0cm 0.0cm, clip]{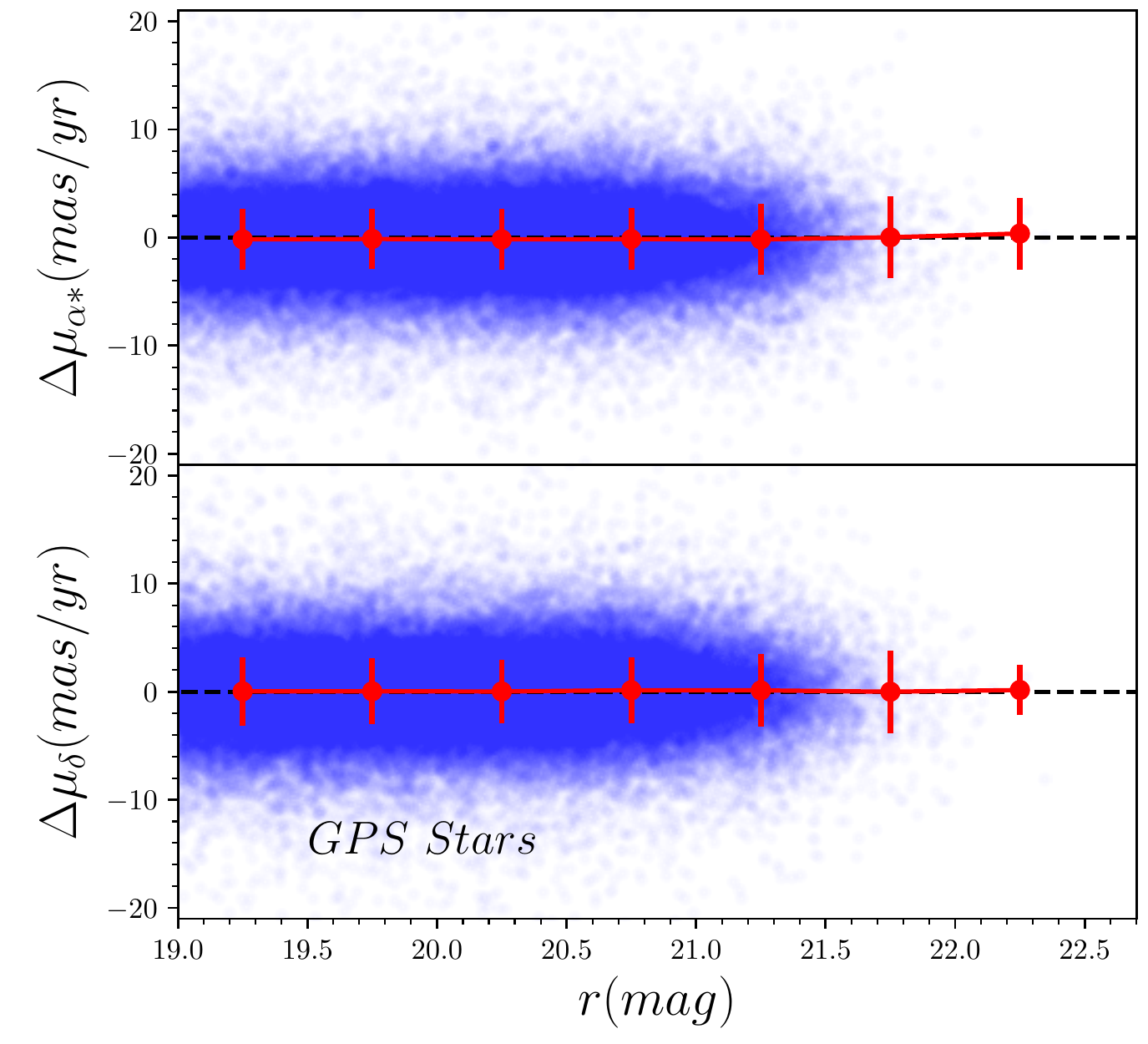}
\caption{Comparison of proper motions between GPS1+ and {\it Gaia} DR2 in different magnitude bins. Two typical proper motion modes are presented: the GP proper motions (the left panel), and the GPS proper motions (the right panel). The blue points are the scatters of the differences of proper motions ($\Delta\mu=\mu_{GPS1+} - \mu_{Gaia}$). The red curves are the median values of $\Delta\mu$ in different magnitude bins and the error bars represent the robust {\it rms}. The black dashed lines mark $\Delta\mu=0$. All the red points oscillate around the black dashed lines  within $\pm$0.05 \masyr, indicating that the average accuracy of GPS1+ proper motions is better than 0.05 \masyr. 
The average {\it rms} in the GP case is $\sim$ 5.0 \masyr, which is reduced to $\sim$3.0 \masyr\ in the GPS case.}\label{fig:vali_gaia}
\end{figure*}

Figure \ref{fig:vali_gaia} illustrates the comparison of proper motions between our GPS1+ and {\it Gaia} DR2 for \mura\ (the top sub-panel) and \mudec\ (the bottom sub-panel). Two typical proper motions are presented: the GP proper motions (the left panel), and the GPS proper motions (the right panel). The median of the differences of proper motions ($\Delta\mu=\mu_{GPS1+} - \mu_{Gaia}$) lies within $\pm$0.05 \masyr\ of zero,  implying that the accuracy of GPS1+ proper motion is better than 0.05 \masyr\ for both \mura\ and \mudec. The red bars indicate the average {\it rms} of GPS1+ proper motion is better than 5.0 \masyr\ in the GP mode, and 3.0 \masyr\ in the GPS mode, respectively. Here, we assume that the proper motions are measured well enough in {\it Gaia} DR2.

\begin{figure*}[!t]
\centering
\includegraphics[width=0.45\textwidth, trim=0.0cm 0.0cm 0.0cm 0.0cm, clip]{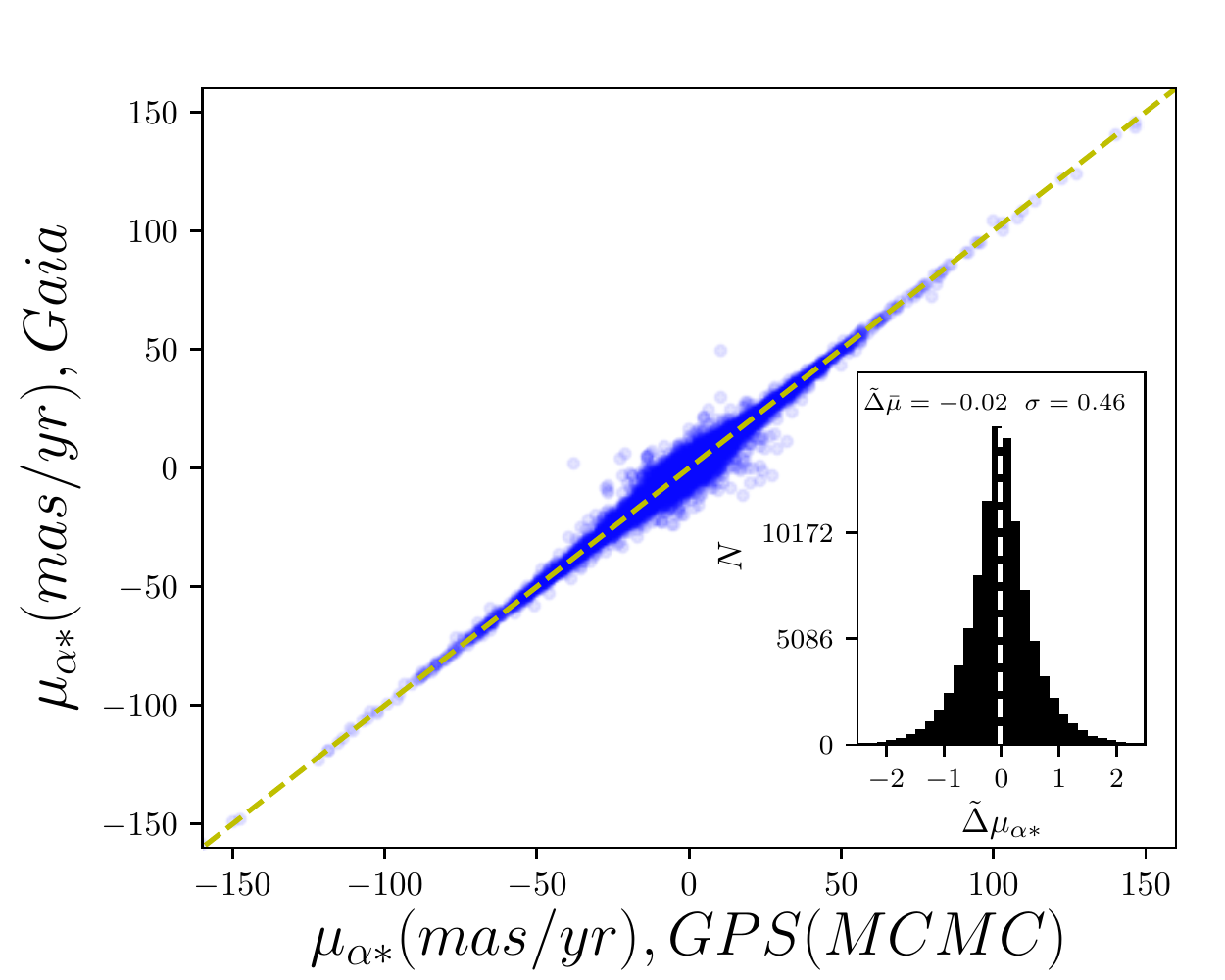}
\includegraphics[width=0.45\textwidth, trim=0.0cm 0.0cm 0.0cm 0.0cm, clip]{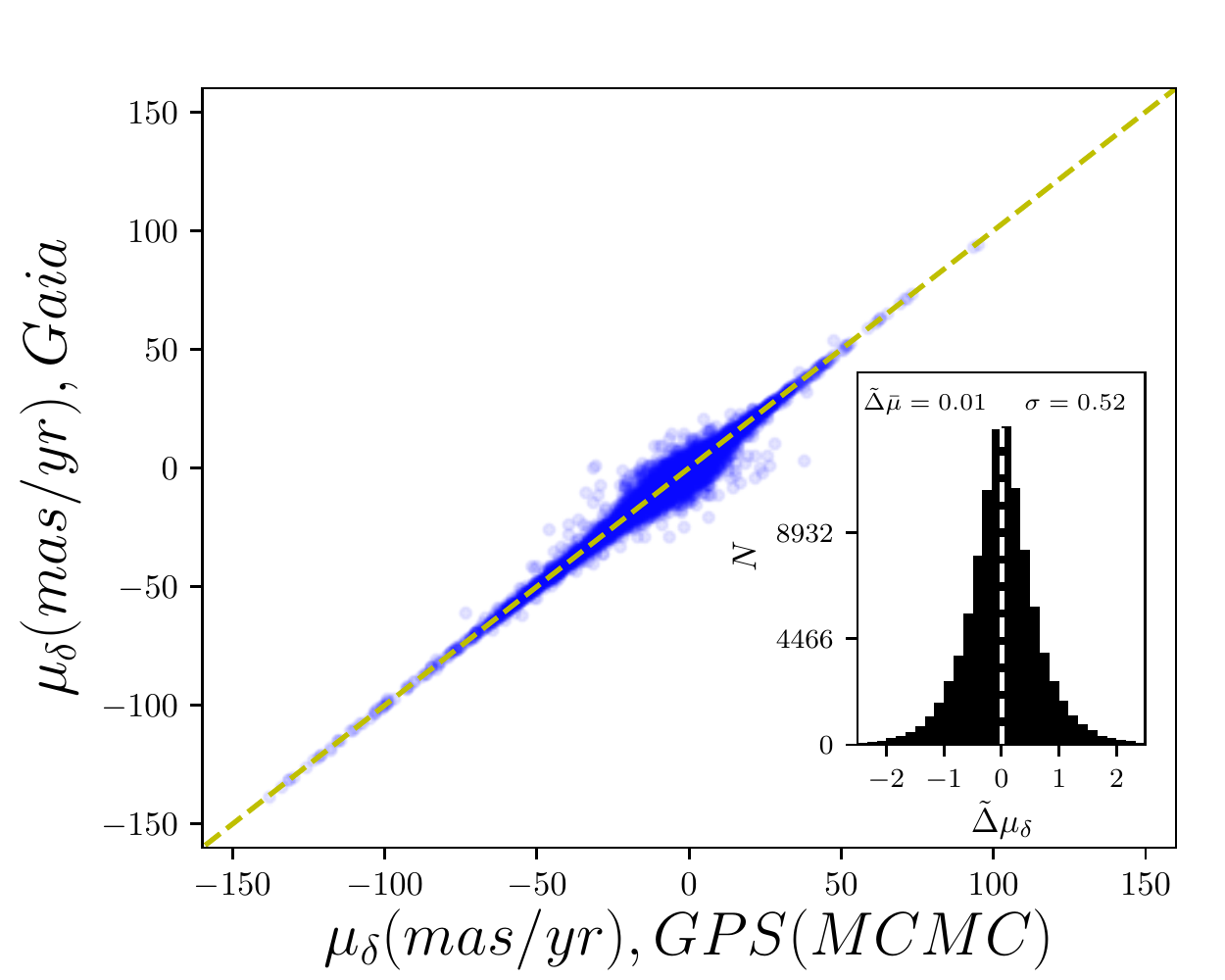}
\caption{Comparison of proper motions between GPS1+ (MCMC) and {\it Gaia} DR2 for \mura\ (the left panel) and \mudec\ (the right panel), based on sources whose proper motions are measured with MCMC fitting. The insets are histograms of the error-weighted difference between our proper motion and {\it Gaia} DR2. The median of the error-weighted differences (the white dashed line) for the \mura\ and \mudec\ are $-0.02\pm0.46$ and $0.01\pm0.52$ (the absolute values: $-0.03\pm0.66$ \masyr and $0.01\pm0.64$ \masyr), respectively.
}
   \label{fig:vali_pal5}
\end{figure*}

Figure \ref{fig:vali_pal5} represents the comparison of proper motions between our GPS1+(MCMC) case and {\it Gaia} DR2 for \mura\ (the left panel) and \mudec\ (the right panel). The insets are the histograms of the error-weighted difference between the two, e,g. $\tilde{\Delta} \mu = (\mu_{ours} - \mu_{Gaia})/\sqrt{\smash[b]{\epsilon_{\mu, ours}^2 + \epsilon_{\mu, Gaia}^2}}$, where the two $\epsilon$ are the errors of our and {\it Gaia} proper motions. The median of the error-weighted differences (marked by the white dashed lines) for the \mura\ and \mudec\ are  $-0.02\pm0.46$ and $0.01\pm0.52$, respectively. The plot indicates that our proper motions are consistent with {\it Gaia} at a high level.

\subsection{Proper Motions Beyond Gaia}\label{sec:val}
In this section, we explicitly summarize what unique data GPS1+ can offer beyond {\it Gaia} DR2. Overall, more than 60\% sources in GPS1+ are beyond the {\it Gaia} limiting magnitude. It means that {\it Gaia} can not reach this part of objects, even in {\it Gaia}'s next data release. The average precision of proper motions for this part of sources is $\sim$7.0\masyr\ if they are measured by SDSS (around one third of them have the astrometry of SDSS). Meanwhile, around 40\% sources are measured new proper motions with the Bayesian technique with the goal of improving the precisions of {\it Gaia} DR2 proper motions at the end faint. Moreover, it is worth to mention that around 13\% sources are the objects whose proper motions are missing in {\it Gaia} DR2. We provide the proper motions for these sources in GPS1+ with an average precision$\sim4.5$\masyr.

\begin{figure}[!t]
\centering
\includegraphics[scale=0.6]{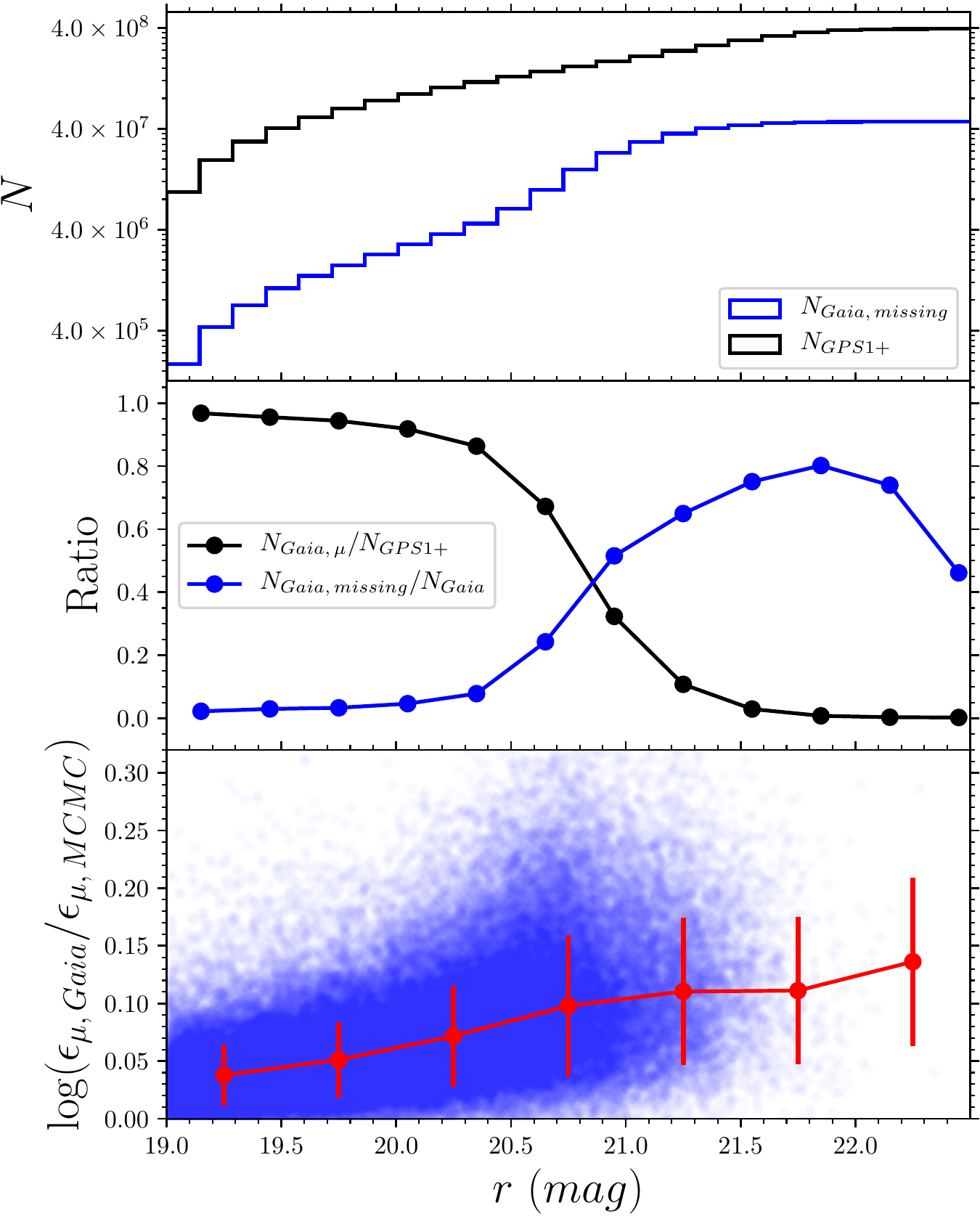}
\caption{Top: the cumulative histograms of the GPS1+ sources ($N_{GPS1+}$, the black curve), and the sources for which GPS1+ provides proper motions, but {\it Gaia} DR2 does not ($N_{Gaia,\,missing}$, the blue curve) across the 3$\pi$ sky over the magnitude at the faint region ($r>19$\,mag). Middle: the number ratio v.s. magnitude. The black and blue curves represent the ratios of $N_{Gaia,\,\mu}/N_{GPS1+}$ and $N_{Gaia,\,missing}/N_{Gaia}$ in the different magnitude bins, where $N_{Gaia,\,\mu}$ and $N_{Gaia}$ are the number of sources for which {\it Gaia} DR2 provides proper motions, and all the sources for which {\it Gaia} DR2 provides positions in a magnitude bin, respectively. Bottom: the precision improvement factor ($\log(\epsilon_{\mu,\,Gaia}/\epsilon_{\mu,\,MCMC})$) of {\it Gaia} DR2 proper motion at the faint region by the Bayesian technique, where $\epsilon_{\mu,\,Gaia}$ and $\epsilon_{\mu,\,MCMC}$ denote the precisions of total proper motions measured in {\it Gaia} DR2 and with the Bayesian technique in GPS1+, respectively. At $r<20.5$\,mag, the precisions of {\it Gaia} DR2 proper motions are tinily improved, only by around 0.05\,dex. At $r>20.5$\,mag, the precisions are improved by about 0.1\,dex on average. Note that this scatter plot is obtained from a sample of one million sources randomly selected from the whole GPS1+ catalog.
}\label{fig:mr_rix}
\end{figure}

Figure \ref{fig:mr_rix} displays the situations of the proper motions beyond {\it Gaia} in the different magnitudes. The top panel illustrates the cumulative histograms of the GPS1+ sources ($N_{GPS1+}$, the black curve), and the sources for which GPS1+ provides proper motions, but {\it Gaia} DR2 does not ($N_{Gaia,\,missing}$, the blue curve), across the 3$\pi$ sky over the magnitude at the faint region ($r>19$\,mag). The two curves tell us the total  $N_{GPS1+}$ and $N_{Gaia,\,missing}$ are about 400 and 47 millions, respectively. The middle panel demonstrates how the ratios of $N_{Gaia,\,\mu}/N_{GPS1+}$ (the black curve) and $N_{Gaia,\,missing}/N_{Gaia}$ (the blue curve) vary with magnitudes. Here, $N_{Gaia,\,\mu}$ and $N_{Gaia}$ are the number of sources for which {\it Gaia} DR2 provides proper motions in GPS1+, and all the sources for which {\it Gaia} DR2 provides positions in a magnitude bin, respectively. The black curve suggests that the number of the sources with  {\it Gaia} proper motions drops dramatically at $r>20$\,mag in GPS1+, and there are almost no {\it Gaia} proper motions beyond $r>21$\,mag. The blue curve demonstrates that the sources whose proper motions are missing in {\it Gaia} DR2 increase quickly at $r>20.5$\,mag. The bottom panel displays how the precisions of {\it Gaia} DR2 proper motions are improved in the different magnitude bins by including PS1 or SDSS astrometry. As the figure shown, the precisions of {\it Gaia} DR2 proper motions are improved with a limited degree, only by around 0.05\,dex at $r<20.5$\,mag. But at $r>20.5$\,mag, the precisions are improved by about 0.1\,dex on average.

\section{The values and Limitations of GPS1+}
For the most part, GPS1+ constitute a catalog that extends the depth of GPS1 from $r<20$\,mag down to 22.5\,mag. It not only fills up some proper motions missed in {\it Gaia} DR2, but also improves the proper motion precision of faint sources in {\it Gaia} DR2. The most important point is that GPS1+ provide new proper motions for a large number of faint sources beyond {\it Gaia} and other existing catalogs. GPS1+ has important values for the studies involved with faints sources, such as precise age of field stars from white dwarf companions \citep[][Qiu et al. in preparing]{Fouesneau2019}, brown \citep{Cook2017, Luhman2018} or ultrcool \citep{Scholz2020} dwarfs,  white dwarf binaries \citep{Parsons2017, Wang2018, Gentile2019,Brown2020,tian2020, Wang2020}, and the sdA problem \citep{Pelisoli2018a, Pelisoli2018b, Pelisoli2019}. Moreover, GPS1+ has some potential values for the studies, such as the stellar kinematics \citep{tian2017b, Farihi2018, WangHF2018, Tian2019}, stellar stream \citep{Fu2018}, hypervelocity Stars \citep{Li2018, Brown2018}, and so on.

In addition, it is worth to summarize  the limitations of GPS1+, and where it should be used with caution:
(1) Some sources may have erroneous proper motions in crowded regions, e.g., nearby globular clusters, partly because blended sources are easily classified erroneously as extended sources during the reference frame is built, and partly because source crowding may lead to systematic errors in source centering.
(2) Some regions are blank in the Galactic plane, particularly in the direction of Galactic center, see Figure \ref{fig:uncertanties_star}. So many sources are included in these regions that our pipeline is hard to process these sources.
(3) Some sources, e.g., QSOs, are significantly affected by the effect of differential chromatic refraction (DCR). {\it Gaia} is a space-based telescope, and its observations are not affected by DCR; while PS1 and SDSS are ground-based telescopes and located in different places, so the two surveys suffer from DCR to a different extent. The combination of different surveys in the proper motion fit may lead to complex DCR effects.
(4) Around one third of sources in GPS1+, i.e., the so-called secondary sub-sample, have an average precision of worse than 15.0 \masyr\ for their proper motions, because most of them are so faint that they are beyond the capability of Gaia's detector, and only have PS1 astrometry. They may have no good applications due to the bad precision.

\section{Conclusions}\label{sect:conclusions}
{\it Gaia} DR2 released proper motions for more than 1.3 billion stars with unprecedented precision in the entire sky region. However, there are some spaces left for the successor of GPS1 proper motion catalog. Firstly, the uncertainties of {\it Gaia} proper motions increase with magnitudes as a function of power law at the faint region ($r>19.0$\,mag), the average uncertainty of {\it Gaia} proper motions become larger than 2 \masyr\ for the sources close to {\it Gaia} limiting magnitude. Secondly, more than 361 million stars have no proper motions, but have positions in {\it Gaia} DR2. Thirdly, about 85\% PS1 sources have no {\it Gaia} proper motions in $21<r<22.5$\,mag, which are beyond {\it Gaia} limiting magnitude. In light of these points, we extend the GPS1 catalog.

With the same procedure of GPS1, we calculated the proper motions for all the PS1 sources fainter than 19\,mag in r-band. For the sources with {\it Gaia} proper motions, we build a Bayesian model by taking {\it Gaia} proper motions as priors to calculate another new proper motion for each source combining all the available astrometry from {\it Gaia} DR2, PS1, SDSS, and 2MASS. Finally, we release the GPS1+ proper motion catalog which contains about 400 million point sources down to 22.5\,mag in r-band, across three quarters of the sky. The systematic error (i.e., accuracy) is $<0.1$ \masyr, but the typical uncertainty (i.e., precision) in the proper motion of a single source is mode-dependent: $\sim$ 14.5\% sources in the GPS1+ catalog are measured proper motions in the GPS mode, the average precision is $\sim$2.0\masyr, $\sim$ 43.6\% and 8\% sources are measured in the GP and PD modes, the precision is $\sim$5\masyr\ on average, but $\sim$ 33.9\% sources are only observed by PS1, the typical precision is worse than 15\masyr. Note that $\sim$13\% sources are the objects whose proper motions are missing in {\it Gaia} DR2, GPS1+ provide their proper motion with a precision of $\sim$4.5\masyr, and $\sim$40\% sources have {\it Gaia} proper motions, we re-calculate their proper motions by building a Bayesian model, the final precision of proper motions can be improved up to $\sim$1.0\masyr\ relative to {\it Gaia}'s values at the faint end.

According to the performance, we divide the GPS1+ catalog into two sub-samples, i.e., the primary sources with a typical precision of 2.0-5.0 \masyr, which have either or both of {\it Gaia} and SDSS astrometry; and the secondary sources with an average precision of worse than 15.0 \masyr, which only have PS1 astrometry. The bad precision makes the secondary sources probably have no good applications.

The GPS1+ proper motions are validated with QSOs, and the performance is illustrated by comparing with proper motions of {\it Gaia} DR2.

\acknowledgements
H.-J.T. acknowledges the National Natural Science Foundation of China (NSFC) under grants 11873034, U1731108, and U1731124. H.-W.R. acknowledges funding from the European Research Council under the European Unions Seventh Framework Programme (FP 7) ERC Grant Agreement n. [321035].The Pan-STARRS1 Survey (PS1) has been made possible through contributions of the Institute for Astronomy at the University of Hawaii, Pan-STARRS Project Office, Max-Planck Society and its participating institutes, specifically  Max Planck Institute for Astronomy, Heidelberg and Max Planck Institute for Extraterrestrial Physics, Garching, Johns Hopkins University, Durham University, University of Edinburgh, Queen's University Belfast, Harvard-Smithsonian Center for Astrophysics, Las Cumbres Observatory Global Telescope Network Incorporated, National Central University of Taiwan, Space Telescope Science Institute, National Aeronautics and Space Administration under Grant No. NNX08AR22G issued through the Planetary Science Division of the NASA Science Mission Directorate, the National Science Foundation under Grant No. AST-1238877, University of Maryland, Eotvos Lorand University and Los Alamos National Laboratory. This work has made use of data from the European Space Agency (ESA)
mission {\it Gaia} (\url{https://www.cosmos.esa.int/gaia}), processed by
the {\it Gaia} Data Processing and Analysis Consortium (DPAC,
\url{https://www.cosmos.esa.int/web/gaia/dpac/consortium}). Funding
for the DPAC has been provided by national institutions, in particular
the institutions participating in the {\it Gaia} Multilateral Agreement.

\bibliographystyle{apj}

\bibliographystyle{yahapj}

\end{document}